\documentstyle[epsf, rotate]{l-aa}
\begin{document}
\thesaurus{(02.01.2; 08.14.1; 09.07.1; 10.07.1; 13.25.3)}
\title {Spatial distribution of the accretion luminosity of
isolated neutron stars and black holes in the Galaxy}

\author{Sergei B.Popov  \and
  Mikhail E.Prokhorov
}

\date{}

\institute{
Sternberg Astronomical Institute,
 Universitetskii pr.13, 119899, Moscow, Russia
}

\maketitle

\markboth{S.B. Popov \& M.E. Prokhorov: Spatial distribution of the
accretion luminosity}{}

\begin{abstract}
    We present here a computer model of the
distribution of the luminosity,
produced by old isolated neutron stars and black holes
accreting from the interstellar medium in the Galaxy in $R$--$Z$ plane.
We show that the luminosity distributions
have a toroidal structure, with the maximum at $\approx 5-6 kpc$.
\end{abstract}


\section{Introduction}

 Old isolated neutron stars (NS) and black holes (BH)
form a big populations of the galactic  sources (about $10^8$--$10^9$
objects in the Galaxy), but most of them are unobserved today.
Less than $10^3$ of young NS appear as
radiopulsars, and no isolated BH are observed at the present moment
(probably, some of them were detected, for example, in the
$ROSAT$ survey, but no one is identified).
As in this article we will speak only about isolated compact objects,
we will not use that adjective: ``isolated'' for NS and BH.

During the last several years, the spatial distribution
and other properties of NS became of great interest, because
as it has been proposed, NS can be observed by the {\it ROSAT}
sattelite in soft X-rays due to accretion from the interstellar medium (ISM)
(see, for example, Treves \& Colpi 1991),
and several sources of this type have been observed (Walter et al. 1996).
BH also can appear as such X-ray sources
(Heckler \& Kolb 1996) with some differences in spectrum and
temporal behaivour (absence of pulsations, for example).
That's why here we try to
obtain a picture of the distribution of the accretion luminosity of
these sources.

Fast rotation and/or strong magnetic field can prevent accretion
onto the surface of the NS. In this case the X-ray luminosity will be
very low (except the case of the transient source formation due to
the envelope around the NS)
(see Popov 1994 and Lipunov \& Popov 1995).
Here we consider only accreting NS.
Most of NS are on the stage of accretion, because their magneto-rotational
evolution usually finishes at this stage approximately
 $10^8$ years after  their birth.
The NS properties (periods etc) on the stage of accretion  depend
upon the magnetic field decay (see Konenkov \& Popov 1997).
BH, of course, can appear only as Accretors.

    In the articles of Gurevich et al. (1993),
Postnov \& Prokhorov  (1993, 1994) it was shown
that NS in the Galaxy form a toroidal structure.
The distribution  of the ISM (see, for example,
Bochkarev 1993) also has the toroidal structure.
The maxima of both distributions roughly coincide.

Therefore, most part of NS (and, as one can say,
BH) is located in the dence regions of the ISM.
Thus the accretion luminosity in these regions should be higher.
The results of  computer simulations of this situation are
presented in this paper.

The  trajectories of NS and BH were computed directly
for specified initial velocity distribution,
the Galaxy gravitational potential and
the distribution of the ISM density.
Preliminary results of such computations for NS for
$\delta$--{\it function} and maxwellian velocity distributions were
presented in Popov \& Prokhorov (1996, paper I).

In the section 2 we briefly describe our model. In the third section
the results and a short discussion are presented. And finaly we
conclued our results in the last section.

\section{The Model}

We solved numerically the system of differential equations of motions
in the given potential.

We used the  Galactic potential, taken in the form (Paczynski 1990):

$$
    \Phi_i (R,Z)=GM_i/\left(R^2+[a_i+(Z^2+b_i^2)^{1/2}]^2\right)^{1/2}
$$
with a quasi-spherical halo with the density distribution in the form:
$$
   \rho=\frac{\rho_0}{1+(d/d_0)}, \,\,\, d^2=R^2+Z^2.
$$

Here $R$ and $Z$ are cylindrical coordinates, $d$-- radial distance in
the quasi-spherical halo.
The parameters of the potential are given in the table,
$\rho_0$ is determined through the halo mass, $M_0$.

\vskip 0.1cm

\noindent
\begin{tabular}{lccc}
Disk &$a_D$=0      &$b_D$=277 pc&$M_D=1.12\cdot 10^{10} M_{\odot}$\\
Bulge&$a_B$=3.7 kpc&$b_B$=200 pc&$M_B=8.07\cdot 10^{10} M_{\odot}$\\
Halo &             &$d_0$=277 pc&$M_0=5.0 \cdot 10^{10} M_{\odot}$\\
\end{tabular}

\vskip 0.1cm

 The density in our model was constant in time. Local density was calculated
using data and formulae from Bochkarev (1993) and Zane et al. (1995).
Here $R$ and $Z$ are cylindrical coordinates (it is assumed, that the Galaxy
has cylindrical symmetry), $n$ -- total concentration,
$n_{HI}$ and $n_{H_2}$ are concentrations of the neutral and molecular
hydrogene, $n_0(R)$, $n_2(R)$ and $n_3(R)$ are the values
of concentrations for $Z=0$.

$$
   n(R, Z)= n_{HI}+2\cdot n_{H_2}
$$

$$
   n_{H_2}=n_2(R)\cdot exp \left[ \frac{ -Z^2}{2\cdot (70 pc)^2} \right]
$$

   If $0kpc \le R \le 3.4 kpc $, then

$$
   n_{HI}=n_0(R)\cdot exp \left[ \frac{-Z^2}{2\cdot (140 pc \cdot R/ 2
kpc)^2} \right], $$

    For $0 kpc \le R \le 2 kpc \, \,  n_0(R) $ was  assumed to be
constant for all $R$:

$$
   n_0(R<2 kpc)=n(R=2 kpc)
$$

    Of course, it is not accurate for small R, so for the very central part
of the Galaxy our results are only a rough estimation (see Zane et al.
(1996) for detailed calculation of NS emiision from the Galactic
center region).   If $3.4 kpc \le R \le 8.5 kpc$, then  $$
   n_{HI}=0.345\cdot exp \left[ \frac{-Z^2}{2\cdot (212 pc)^2} \right] +
$$
$$
          0.107\cdot exp \left[ \frac{-Z^2}{2\cdot (530 pc)^2} \right] +
$$
$$
          0.064\cdot exp \left[ \frac{-Z}{403 pc} \right]
$$

    If $ 8.5 \le R \le 16 kpc$, then

$$
   n_{HI}=n_3(R)\cdot exp \left[ \frac{-Z^2}{2\cdot (530 pc
\cdot R / 8.5 kpc)^2} \right]
$$
$n_0(R)$, $n_2$ and $n_3(R)$ were taken from Bochkarev (1993).

    The density distribution in the $R$-$Z$ plane
used in our computations is shown in the figure 1.

\begin{figure}
\epsfxsize=8cm
\centerline{{\epsfbox{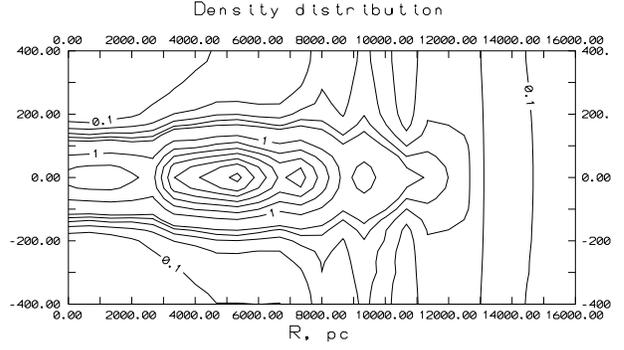}}}
\caption{ The density distribution in units particle per
cubic centimeter in the $R$--$Z$ plane.}
\end{figure}

 In our model we assumed, that the birthrate of NS and BH is
proportional to the square of the local density.
Stars were born in the Galactic plane (Z=0) with circular velocities
plus additional isotropic kick velocities.

For the kick velocity we used the formula from Lipunov et al. (1996).
It was constructed as an analitical approximation of the
three-dimensial velocity distribution of radiopulsars from
Lyne \& Lorimer (1994).

$$
  f_{LL}(V)\propto \frac{x^{0.19}}{(1+x^{6.72})^{1/2}},
$$
here $V$-- spatial velocity of the compact object,
$V_{char}$ -- characteristic velocity, $x=V/V_{char}$,
$f_{LL}$-- the probability (see
the detail describtion of the analitical approximation in  Lipunov et
al. (1996)). This formula reproduces the observed distribution
with the mean velocity 350 km/s for $V_{char}$=400 km/s.
This velocity distribution seems more luckely, than the
$\delta$--{\it function} and Maxwellian distributions, which we
used in Paper I. Kick velocities were taken equal for NS and BH
(this is a ``zero-hypothesis'', as we don't have any exact
information on this matter). But it is possible, that BH have lower
kick velocities because of their high masses.
One of the reasons to make computations for $V_{char}$=200 km/s
was to explore this situation.

For each star we computeded the exact trajectory and the
accretion luminosity.
The accretion luminosity was calculated using Bondi formula:

$$
    L=\left(\frac{GM \dot M}{R_{lib}}\right)
$$

$$
   \dot M=2 \pi \left(
    \frac{(GM)^2 \rho(R,Z)}{(V_s^2+V^2)^{3/2}}\right)
$$.

    Sound velocity,$ V_s$, was taken to be 10 km/s everywhere.
We used equal masses $M_{NS}=1.4 M_{\odot}$ for NS and
$M_{BH}=10 M_{\odot}$
for BH. Density is determined as:
$\rho =n \cdot m_H$, where $m_H$-- the mass of the hydrogen atom. Radii,
$R_{lib}$, where the energy is libirated, were assumed to be equal to
10 km for NS and 90 km (i.e. $3\cdot R_g$, $R_g=2GM/c^2$)
for BH.  Statistics was collected on
the grid with the cell size 100 pc in R-direction and 10 pc in
Z-direction (centered at R=50 pc, Z=5 pc and so on).  The luminosity is
shown on the graphs in ergs per second per cubic parsec.

For the normalization of our results we used an assumption, that there are
$N_{NS}=10^9$ NS and $N_{BH}=10^8$ BH in the calculated volume
of the Galaxy. For Salpeter mass function with $\alpha$=2.35
the ratio of NS to BH is about 10 if NS are formed from the
stars with masses between $10M_{\odot}$ and $\approx 45-50 M_{\odot}$,
and BH from the stars with masses higher than $\approx 45-50
M_{\odot}$.  Motch et al. (1997) argued, that $N_{ns}=10^9$ can be
ruled out, and $N_{NS}=10^8$ is a more probable value, but for the
caclculations of the distribution the total  number is not so
improtant, and for other numbers of compact objects
the results (i.e. the value of the luminosity) can be easily scaled.

\section{Results and discussion}

    On the figures 2--5 we represent the results for two characteristic
values of the velocity distribution for NS and BH.
On the graphs 2--5
the region $16000 \, pc\, \times \, 800\,  pc$ for averaged
accretion luminosity for the azimutale angle 0--180 degrees is shown.
We  show only this region, because we are mainly
interested in the part, where the maximum is situated, but while
the computations NS and BH could move inside infinite region.
The scale for $R$ and $Z$ axes are different in order to show clearly the
structure in $Z$ direction. Differences between the luminosity
distribution for $Z>0$ and $Z<0$ demonstrate accuracy of the
computations (curves were not smoothed).

\begin{figure}
\epsfxsize=8cm
\centerline{{\epsfbox{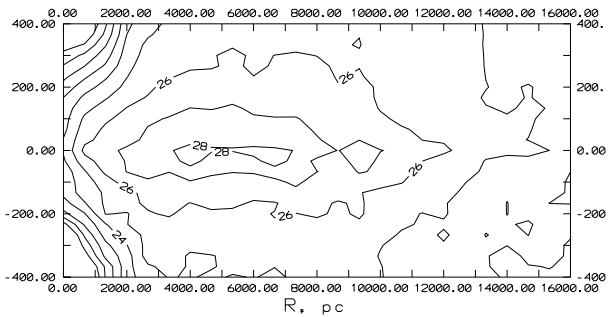}}}
\caption{ The accretion luminosity distribution in $R$--$Z$ plane for
neutron stars for characteristic kick velocity 200 km/s.
The luminosity is shown in ergs per second per cubic parsec.
$N_{NS}=10^9$}
\end{figure}

\begin{figure}
\epsfxsize=8cm
\centerline{{\epsfbox{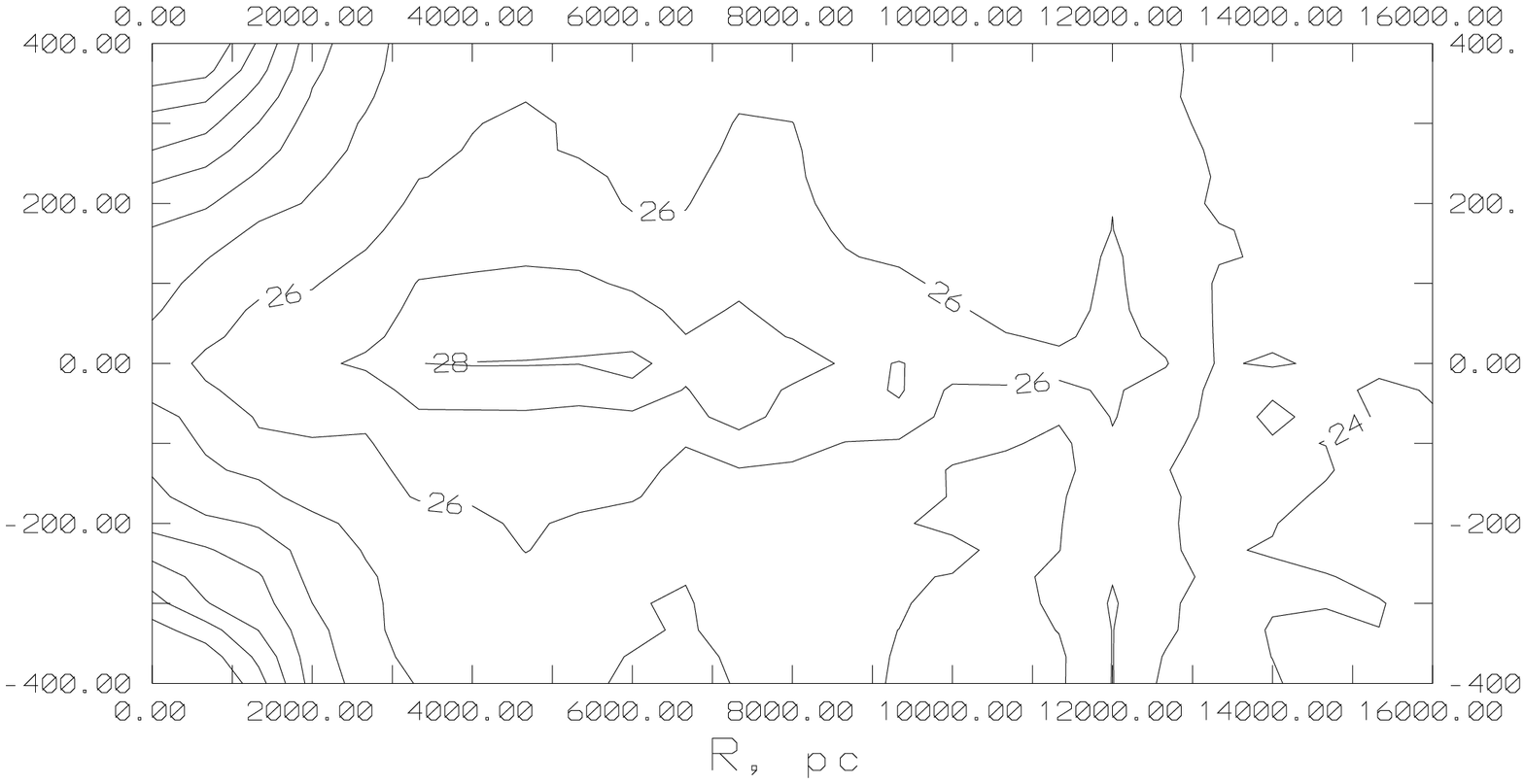}}}
\caption{ The accretion luminosity distribution in the $R$--$Z$ plane for
neutron stars for characteristic  kick velocity 400 km/s.
The luminosity is shown in ergs per second per cubic parsec.
$N_{NS}=10^9$}
\end{figure}

\begin{figure}
\epsfxsize=8cm
\centerline{{\epsfbox{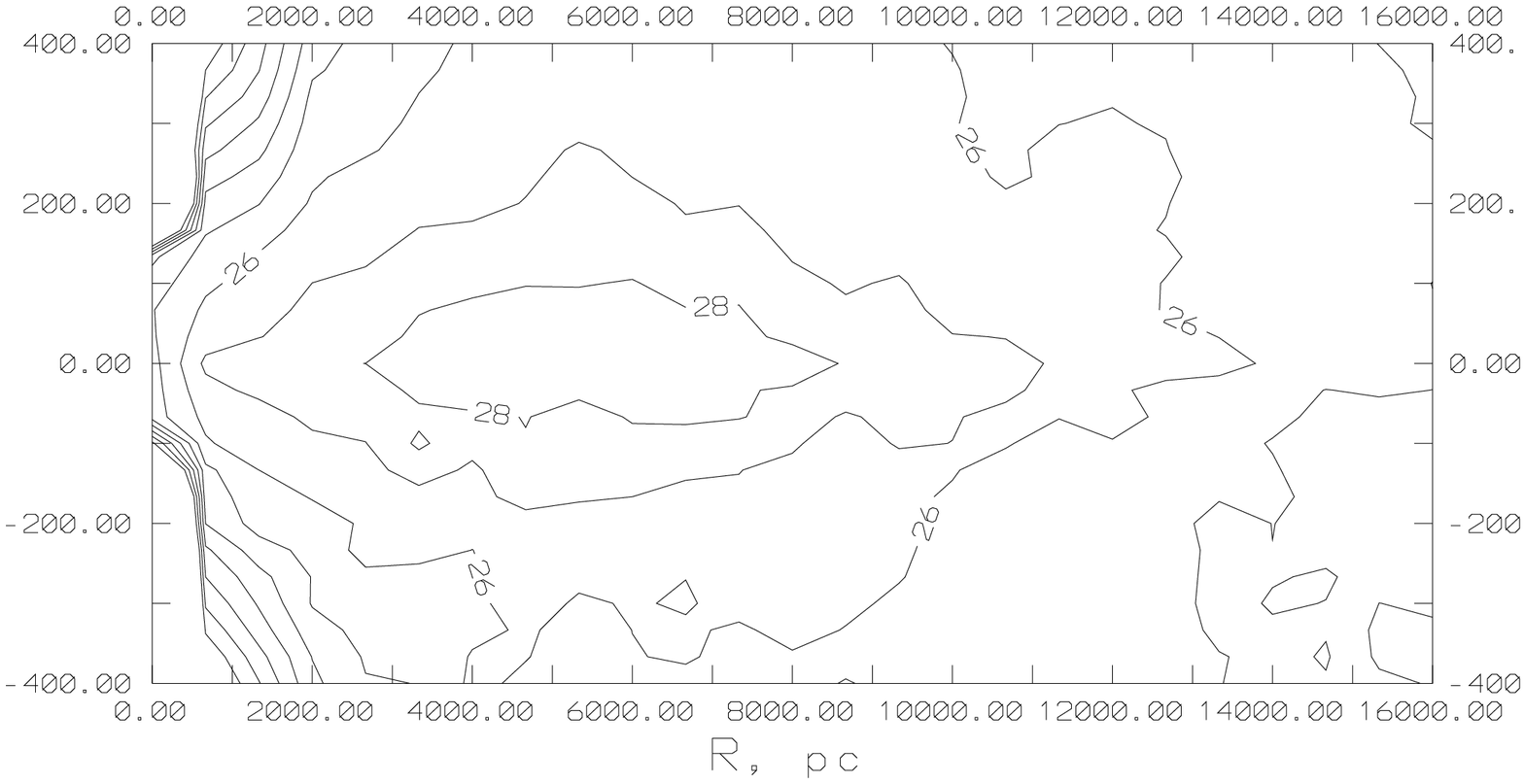}}}
\caption{The  accretion luminosity distribution in the $R$--$Z$ plane for
black holes for characteristic  kick velocity 200 km/s.
The luminosity is shown in ergs per second per cubic parsec.
$N_{BH}=10^8$}
\end{figure}

\begin{figure}
\epsfxsize=8cm
\centerline{{\epsfbox{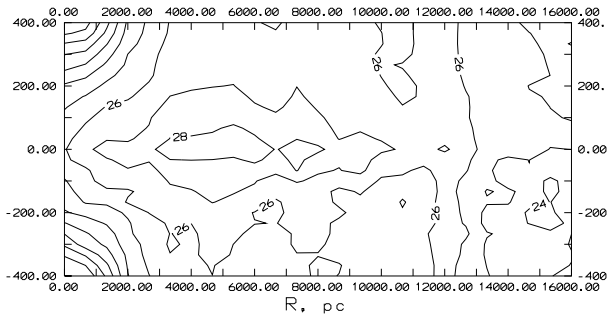}}}
\caption{The  accretion luminosity distribution in the $R$--$Z$ plane for
black holes for characteristic  kick velocity 400 km/s.
The luminosity is shown in ergs per second per cubic parsec.
$N_{BH}=10^8$}
\end{figure}

On the figure 6 the slice at Z=+5 pc (the first grid cells
in the positive Z-direction)
for the characteristic  kick  velocity $V_{char}$= 200 km/s is shown:
luminosity  vs. radius.
The figure is not symmetric. The right part corresponds to the
azimutale angles 0--180 degrees, the left - 180--360 degrees.
Differences between the left and the right parts of the curve
demonstrate how precise our method is.

\begin{figure}
\epsfxsize=7cm
\centerline{\rotate[r]{\epsfbox{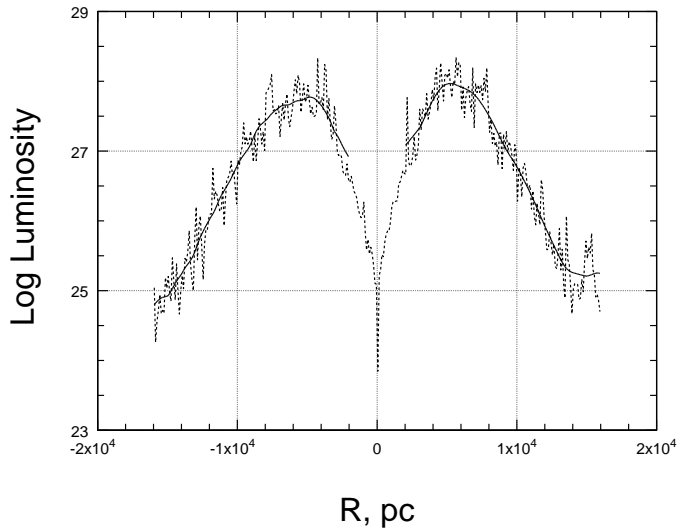}}}
\caption{Slice at Z=+5 pc for NS for the characteristic  kick velocity
200 km/s. $N_{NS}=10^9$. The accretion  luminosity is shown  in ergs per second
per cubic parsec.  With the solid line we show the smoothed curve.  }
\end{figure}

\section{Concluding remarks}

    As it is clearly seen from the figures, the distribution of
the accretion luminosity  in R-Z plane
forms a toroidal structure with the maximum
at approximatelly 5 kpc.

 As expected, for BH we have relatively higher luminosity, than for NS,
as BH have greater masses.
But if the total number of BH is significantly lower than the number
of NS, their contribution
to the luminosity can be less than the contribution of NS.
The total accretion lumiunosity of the Galaxy for $N_{NS}=10^9$
and $N_{BH}=10^8$ is about $10^{39}-10^{40}${\it erg/s}.
For the characteristic velocity 200 km/s for NS the maximum of the
distribution is situated approximately at  5.0 kpc  and for BH at  4.8
kpc.  For NS with the characteristic velocity 400 km/s maximum is
situated at 5.4 kpc, and for BH at 5.0 kpc.  This result is also
expected because of the same reason (higher masses of the BH with the
same value of kick velocity).

    The toroidal structure of the luminosity distribution
of NS and BH is an interesting and important feature
of the Galactic potential.
As one can suppose, for low characteristic kick velocities and for BH
we obtained higher luminosity.

    As me made very general assumptions, we argue, that such a distribution
is not unique only for our Galaxy, and all spiral galaxies can have such
a distribution of the accretion luminosity, associated with accreting NS
and BH.

\begin{acknowledgements}
We thank I.E. Panchenko for his help in the English translation.
    The work was supported by the RFFI (95-02-6053) and
the INTAS (93-3364) grants.
The work of S.P. was also supported by the ISSEP.
\end{acknowledgements}

\end{document}